\documentclass{ws-procs9x6}

\begin{document}

\title{THREE-JET ANGULAR CORRELATIONS AND SUBJET DISTRIBUTIONS AT ZEUS}

\author{T.~SCH\"ORNER-SADENIUS\footnote{
    \uppercase{T}alk given on
    behalf of the \uppercase{ZEUS} collaboration at \uppercase{DIS}06,
    \uppercase{T}sukuba, \uppercase{J}apan, \uppercase{A}ril 2006.}
}

\address{Hamburg University, IExpPh,\\
Luruper Chaussee 149, 22761, Germany,\\  
E-mail: schorner@mail.desy.de}

\maketitle

\abstracts{
Besides structure function measurements and jet physics, there 
is a lively collection of more specific QCD analyses at HERA. In this
contribution we 
present three-jet angular correlations and subjet distributions measured in ep 
collisions with the ZEUS detector. The angular correlations provide
sensitivity to the color factors of the underlying gauge group and thus
facilitate tests of basic properties of the strong interaction. 
The subjet distributions allow
tests of the QCD radiation pattern within a jet in the perturbative regime. 
}

\section{Three-Jet Angular Correlations}

QCD is widely accepted as the theory of the strong interaction. Nevertheless,
it is worthwhile to test basic properties of QCD. In two new 
analyses\cite{angular} ZEUS
investigated the color factors of QCD which define the relative strengths of
the various QCD vertices. 

In three-jet production, various combinations of color factors contribute to
the cross-section which, in leading order, symbolically can be written as: 
\mbox{$\sigma_{3jet}=C_F^2\cdot\sigma_A + C_FC_A\cdot\sigma_B + C_FT_F\cdot\sigma_C + T_FC_A\cdot\sigma_D.$}
Here, $C_F$, $C_A$ and $T_F$ are the color factors of the $q\rightarrow qg$,
the $g\rightarrow gg$ and the $g\rightarrow q\overline{q}$ vertices,
respectively. The $\sigma_i$ denote the contributions to the cross-section for 
the color factor combination in question. Special note should be given to the
contributions containing the color factor $C_A$ since the three-gluon
vertex is a very specific feature of non-abelian gauge theories such as QCD.  

\begin{figure}[t]
\hspace*{-0.5cm}
\epsfxsize=6.cm   
\epsfbox{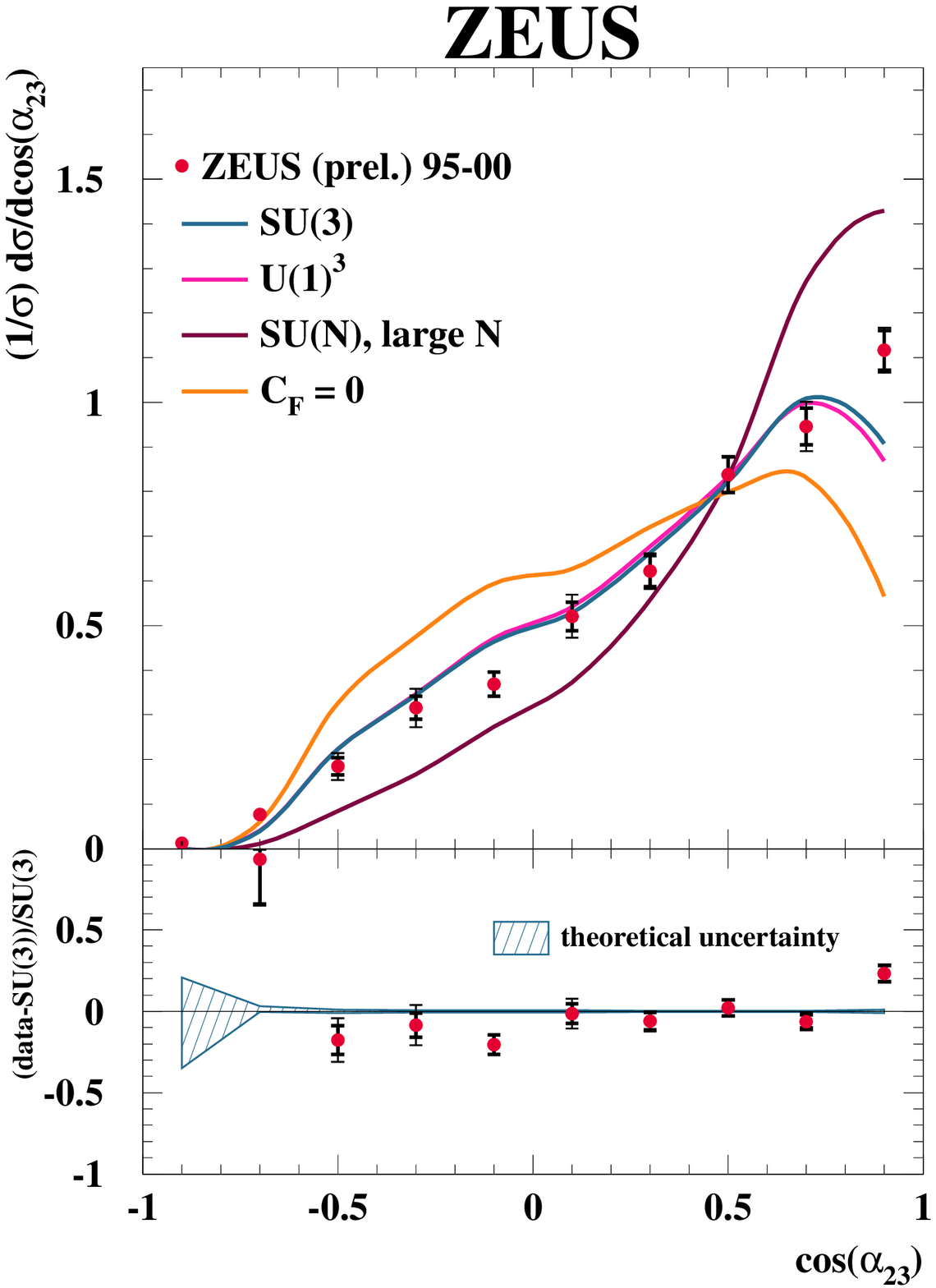}
\hspace*{-1.cm}
\epsfxsize=6.cm   
\epsfbox{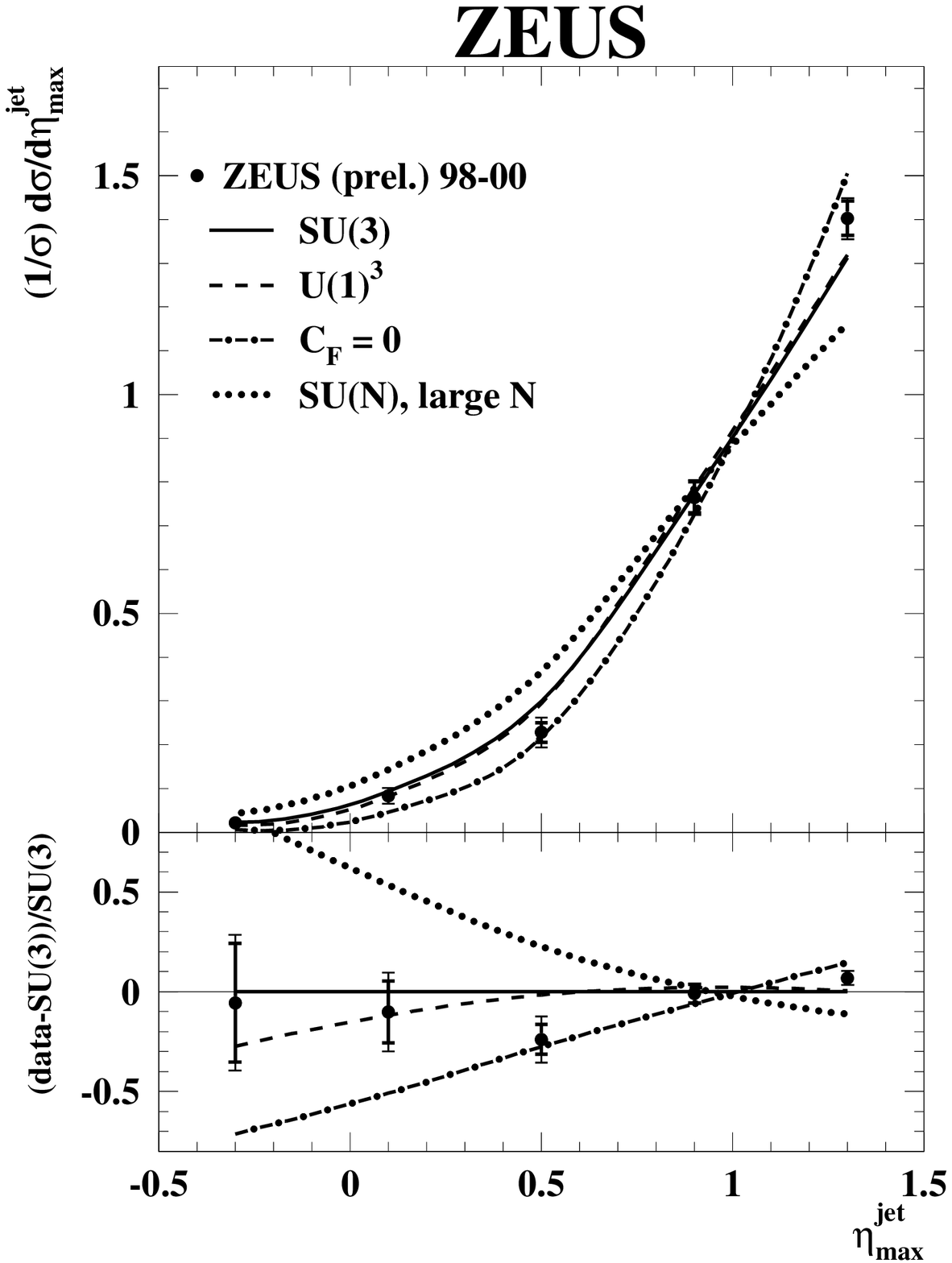}
\caption{Left: normalised cross-section d$\sigma$/d$\cos\alpha_{23}$ for the
  photoproduction analysis. Right: normalised cross-section 
  d$\sigma$/d$\eta_{max}^{jet}$ for the
  DIS analysis. \label{figure1}}
\end{figure}

In DIS, the analysis of the three-jet correlations is restricted to 
81.7~${\rm pb^{-1}}$ from the years 1998-2000 with 
$Q^2 >$~125~${\rm GeV^2}$. Three jets 
with transverse energies (in the Breit
frame) of at least 8, 5 and 5~GeV had to be reconstructed
using the longitudinally invariant $k_T$ cluster algorithm; 
the jets had to be in the pseudorapidity interval
\mbox{-2~$<\eta_{Breit}<$~1.5}.  

In photoproduction, 127~${\rm pb^{-1}}$ from
1995-2000 were analysed; the three jets were all
required to have at least 14~GeV transverse energy and to be well contained
in the detector acceptance, -1~$< \eta_{lab} <$~2.5. In addition, the
photoproduction analysis was restricted to a data sample enriched in
direct photon-parton interactions using the
quantity $x_{\gamma}^{obs}$, $x_{\gamma}^{obs} >$~0.7. 

Normalised cross-sections were measured and compared to both
leading-order MC models and to fixed-order QCD calculations for a
number of 
observables. Examples are $\Theta_H$, the angle between the plane determined
by the highest transverse energy jet and the beam and the plane determined by
the two lowest transverse energy jets, or the cosine of the angle between the
two lowest transverse energy jets, $\cos\alpha_{23}$.
In DIS, also the pseudorapidity of the most forward jet in the Breit frame,
$\eta_{max}^{jet}$, was measured.
In both the DIS and the photoproduction analysis the fixed-order calculations
were only at leading order; they nevertheless provided access to the color
factors and thus allowed to change the gauge group underlying the calculations.

Figure~\ref{figure1} (left) shows the normalised three-jet
cross-section as a function of the observable $\cos\alpha_{23}$ for the
photoproduction analysis. The data are compared to a fixed-order calculation
with four different settings of the color factors, one of which corresponds to
the QCD gauge group SU(3). Also shown is an abelian gauge group, ${\rm
  U(1)^3}$, which is similar to QCD except for the triple-gluon vertex. As can
be seen, the two other, rather extreme choices of color factors are
excluded by the data, but there is little discrimination power between SU(3)
and the abelian model. The same statement holds also for the other observables
under study.  

On the right hand side of figure~\ref{figure1}, the normalised cross-section
as function of $\eta_{max}^{jet}$ is shown for the DIS
analysis. The data are compared to the same four different color factor
choices, and again the SU(N) and $C_F=0$ models can clearly be excluded
whereas there is little discrimation between QCD and the abelian model.
New angular correlations need to be designed that enhance the contribution from
the triple-gluon vertex to discriminate between SU(3) and ${\rm U(1)^3}$.

\section{Subjet Distributions}


At high transverse energies, when jet fragmentation effects
become negligible, jet structure can be described perturbatively. The lowest
non-trivial (LO) contribution to the jet structure is given by
$\mathcal{O}(\alpha_S)$ pQCD calculations in the laboratory
frame with one or two partons in one jet. 
Next-to-leading order (NLO) calculations in this frame are feasible since it
is possible to have up to three partons in one jet. In a new measurement\cite{subjets}
the internal structure of jets is analysed in terms of subjets. 
Subjets within a given jet identified by the $k_T$ cluster
algorithm are identified by re-applying the algorithm to all particles of a jet
and clustering until for all particle pairs $i,j$ the quantity 
$d_{ij}=min\left( E_{T,i},E_{T,j}\right)\cdot\left( \Delta\phi^2_{ij}+\Delta\eta^2_{ij}\right)$ is
greater than $d_{cut}=y_{cut}\cdot E_T^2$. $E_{T,i}$ is the transverse
energy of particle $i$, and $\Delta\phi_{ij}$ ($\Delta\eta_{ij}$) 
is the difference in
azimuthal angle (pseudorapidity) of particles $i$ and $j$. 

\begin{figure}[t]
\hspace*{-0.5cm}
\epsfxsize=6.cm   
\epsfbox{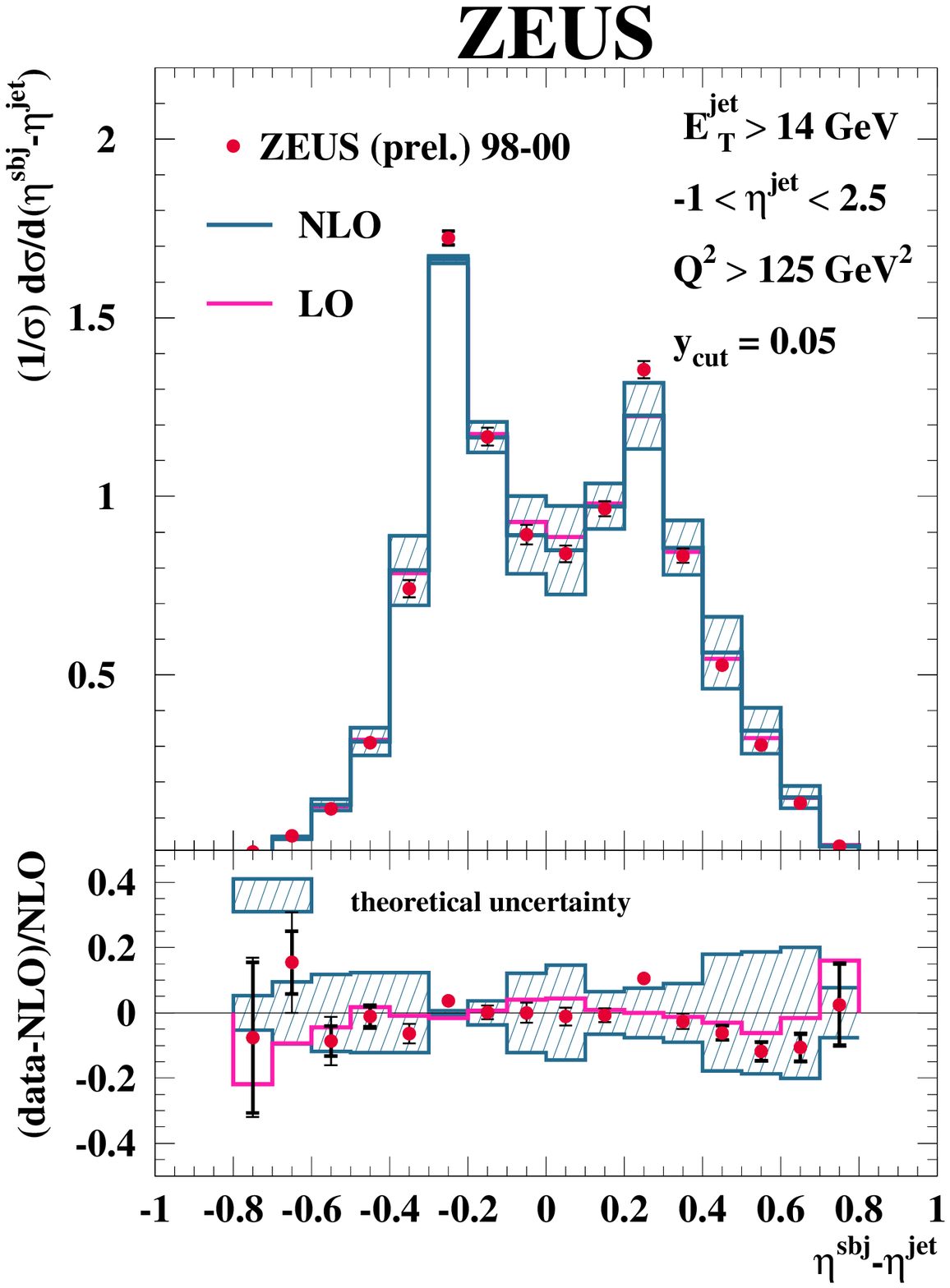}
\hspace*{-1.cm}
\epsfxsize=6.cm   
\epsfbox{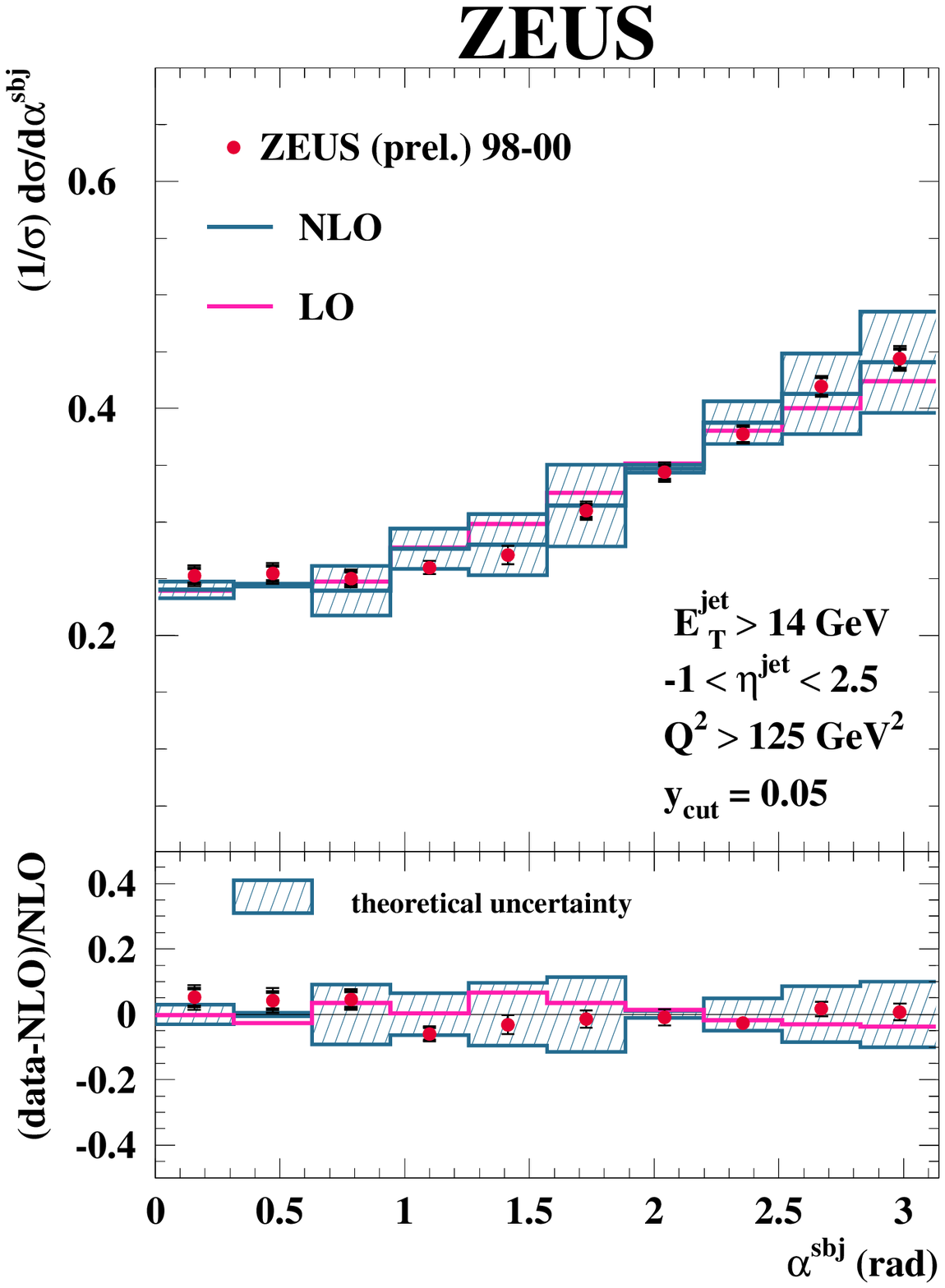}
\caption{Left: normalised subjet cross-section as function of the
  pseudorapidity difference $\eta^{sbj}-\eta^{jet}$. 
  Right: normalised subjet cross-section 
  as function of $\alpha^{sbj}$.
  \label{figure2}}
\end{figure}


Subjet cross-sections are measured in 81.7~${\rm pb^{-1}}$ of ZEUS data collected
in 1998-2000. The kinematic range of the analysis is
restricted to $Q^2 >$~125~${\rm GeV^2}$. Jets were reconstructed in the
laboratory frame using the $k_T$ cluster algorithm. Subjets were then
reconstructed in jets 
with transverse energies of at least 14~GeV in the pseudorapidity range 
-1~$< \eta_{lab} <$~2.5. The final sample consisted of jets
with exactly two subjets $y_{cut} =$~0.05.

Subjet cross-sections were measured as functions of the difference in
transverse energy between the subjet and the jet, of the difference in
azimuthal angle or in pseudorapidity, and of 
the angle between the highest transverse energy subjet and the beam
line in the pseudorapidity-azimuth plane, $\alpha^{sbj}$. 
The measured distributions were
compared to leading order MC models, resulting in a good
description of the data, and to fixed-order QCD calculations with up
to three partons in one jet.    


Figure~\ref{figure2} (left) shows the normalised subjet
cross-section as a function of the difference in pseudorapidity between subjet
and jet. The data are well described by both LO and NLO 
calculations and show that the
highest-energy subjet tends towards the rear direction. The same behaviour is
observed for
$\alpha^{sbj}$ which is shown on the right hand side of
figure~\ref{figure2}. Also this distribution is well described by the QCD
calculations. 

\vspace*{-.1cm}

\section{Conclusion}

An investigation of the color factors of the strong interaction in three-jet
correlations in photoproduction and DIS can exclude some 
exotic candidates for the gauge structure of the theory of the strong
interaction. However, QCD/SU(3) and an abelian model of type ${\rm U(1)^3}$
cannot be separated with the angular correlations studied. 

Studies of subjet cross-sections in DIS show
that the pattern of parton radiation within jets in the perturbative regime can be
described by fixed-order QCD calculations with up to three partons in a jet.

\end{document}